\documentclass[useAMS,usenatbib,usegraphicx]{mn2e}
\usepackage{amsmath}
\title[Reverberation in RE J1034+396]{X-ray Reverberation close to the black hole in \mbox{RE J1034+396}}
\author[Zoghbi et. al]{A. Zoghbi\thanks{E-mail:
azoghbi@ast.cam.ac.uk} and A. C. Fabian\\
Institute of Astronomy, Madingley Road, Cambridge CB3 0HA}
\begin{document}

\date{}

\pagerange{\pageref{firstpage}--\pageref{lastpage}} \pubyear{2011}

\maketitle

\label{firstpage}

\begin{abstract}
In previous work, we discussed the detection of reverberation delays in the NLS1 1H0707-495. The delays originate close to the black hole. Here, we show that \mbox{RE J1034+396} shows very similar lag properties. At low frequencies ($<1\times10^{-4}$ Hz), the time lag between energy bands increases with energy separation, similar to that commonly seen in Galactic black holes and other AGN. At higher frequencies ($\sim3.5\times10^{-4}$ Hz), the soft ($<1$ keV) and hard ($>3$ keV) bands lag behind the intermediate band (1--3 keV). The simplest interpretation is that the intermediate band is dominated by the direct power-law continuum, while the soft and hard bands are dominated by the relativistically-smeared reflected emission. The low frequency delays are present in both available observations. The high frequency lags are only seen in one observation. In the observation where high frequency reverberation delays are observed, the spectrum contains a power-law component and there is a QPO in the light curve. In the other observation, no power-law component is required and no QPO is seen. The lags originate a few gravitational radii from the black hole, and the QPO is associated with the power-law emitting corona.

\end{abstract}


\section{Introduction}
RE J1034+396 is a Narrow Line Seyfert 1 galaxy (NLS1, $z=0.042$). It has an unusual multi-wavelength
spectrum, with no significant big blue bump in the UV/Optical bands. It is very bright in the EUV and
soft X-rays making it one of the brightest
objects with a so-called `soft excess' (\citealt{1995MNRAS.276...20P,1995MNRAS.277L...5P}). The
soft excess below $\sim2$ keV is a common feature in the X-ray spectra of many AGN, particularly of the
NLS1 class. It is too hot to be thermal blackbody emission from the accretion disc, while the
similarity of its shape in objects with different masses argues against an origin in a cool
Comptonised region existing along with the hot corona emitting at higher energies. A corona of this
type is expected to depend on the seed photons from the disc, which in turn depend on the mass of
the black hole (\citealt{2004MNRAS.349L...7G}). Atomic processes are more likely to be at work
through line-of-sight absorption or reflection. The smoothness of the observed
emission rules out the former (\citealt{2008MNRAS.386L...1S}), while the latter works if the
emission originates very close to the black hole where strong relativistic effects acts to smear out
sharp atomic features (\citealt{2006MNRAS.365.1067C,2011MNRAS.410.1251N}).

One other object, 1H0707-495, also shows a strong soft excess below 1 keV. It was shown recently
that because of the high iron abundance, its soft excess is mainly due to a strong iron L
emission (\citealt{2009Natur.459..540F,2010MNRAS.401.2419Z}) produced by partially
ionised reflection (\citealt{2005MNRAS.358..211R,1991MNRAS.249..352G}). The compactness of the
reflecting region and its proximity to the black hole produces strong relativistic broadening that
smears out the reflection features. The iron L-line is observed together with the more commonly seen broad K
line (e.g. see \citealt{2007ARA&A..45..441M} for a review).

1H0707-495 also shows a time delay between the hard band (1.0--4.0 keV) dominated by the
direct power-law emission, and the soft band (0.3--1.0 keV) dominated by the reflected emission
(\citealt{2009Natur.459..540F,2010MNRAS.401.2419Z}). The soft lag is the first to be
detected, and is a confirmation that the soft band is dominated by reprocessed
emission. The magnitude of the lag itself (30--50 seconds) is consistent with light travel time at
$\sim 2$ gravitational radii from the black hole, as inferred from the energy spectrum. Detailed
energy-dependent lag spectra show that the lag traces the shape of the reflection spectrum, further
confirming its interpretation (\citealt{2011MNRAS.412...59Z}).

In this work, we investigate the nature of the soft excess in RE J1034+396 drawing on the similarities with 1H0707-495. We mainly concentrate
on time lags and their interpretation. Spectral fitting and spectral variability properties
have already been investigated in \cite{2009MNRAS.394..250M}. In that work, the increase of rms variability with
energy was the basis of favouring a low-temperature Comptonisation interpretation for the soft
excess. This component remains constant while the variability is produced by the hard tail changing
in normalisation. Although such low-temperature Comptonisation is not generally seen in AGN, they
suggest that RE J1034+396 is accreting at super-Eddington rates. It is therefore exceptional, which could 
also explain the fact that it is the only AGN to show quasi-periodic
oscillations in its light curve.

\section{Observations \& data reduction}
We use two publicly available XMM-\emph{Newton} observations of RE J1034+396. These were taken in
2007 (obs. id 0506440101) and 2009 (obs. id 0561580201). The EPIC-PN detector was used in full
and small window modes in the first and seconds observations respectively. The total exposure after
removing high background periods was about 80 and 50 ks for the first and second observations
respectively. The first observation was piled-up. The spectra were extracted in different ways to test the effect of pileup. First, we used PATTERN=0 events only, and although this may give the wrong total number of counts, the spectral shape is not distorted. Also, we extracted the spectra using different extraction regions where the central highly piled-up parts of the Point Spread Function (PSF) were excluded. In the spectral fitting, spectra extracted in the different ways have minor differences. Although for the purpose of interpreting the lags, we are only interested in a general spectral model, very good fits are found regardless of the type of extraction, but with parameters that differ slightly which is irrelevant for the this work.
We also calculated the lag with and without removing the central regions of the PSF. These are discussed further in Sec. \ref{sec:time_lag}.

\section{Time Lags}
\subsection{Lag measurements}\label{sec:time_lag}
It has been known that lags in black holes depend on both frequency and energy at the same time
(e.g. \citealt{1999ApJ...510..874N} for galactic black holes and
\citealt{2007MNRAS.382..985M,2008MNRAS.388..211A} for AGN). In \cite{2011MNRAS.412...59Z}, we showed that the
high frequency lags in 1H0707-495 (reverberation lags) are closely linked to the energy spectrum.
In order to measure any time delays, the energy bands have to be selected appropriately.  For RE
J1034+396 however, the components of the energy spectrum are not well known. The whole
energy band (0.3--10 keV) and frequency bands need to be explored. The two observations are
so different in terms of their spectra and light curve properties, that we explore them separately.

Light curves were extracted in 50 energy bands with time bins of 200 seconds. These
were selected so that they have roughly equal number of counts. Fourier-resolved time lags
were calculated between each light curve and a light curve constructed from the whole 0.3--10
keV band, excluding the current band. Taking the whole band as a reference instead of a single
individual band maximises the signal to noise ratio, while removing the current band from the reference
ensures the noise remains uncorrelated between the bands (e.g. see \citealt{2011MNRAS.412...59Z,2011arXiv1104.0634U}). Time lags and their errors were calculated
following standard methods (\citealt{1999ApJ...510..874N}). The frequency axis was binned in
log-steps.
\begin{figure}
\centering
 \includegraphics[width=240pt,clip ]{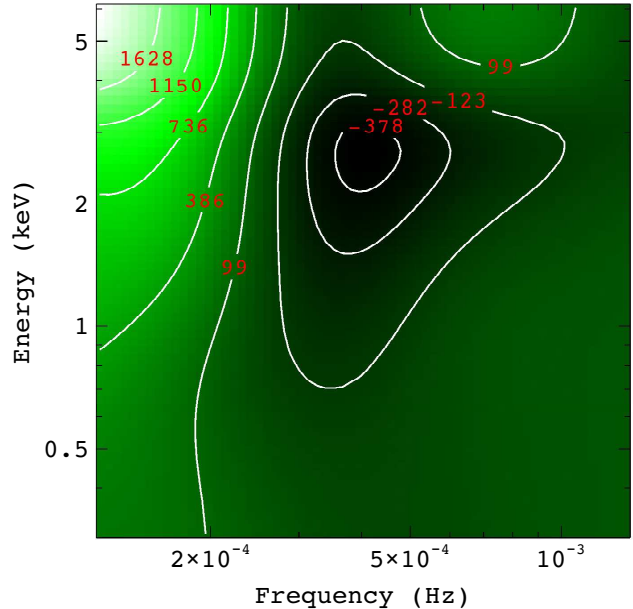}
\caption{2D plot of the lag as a function of energy and frequency. The lags have been smoothed to
help identify the general trends in the lag spectra. The plot is for the first observation only as
the two observations have different lag properties (see text for details).}
\label{fig:en_fq}
\end{figure}

Fig. \ref{fig:en_fq} shows a 2d plot of the lag as a function of energy and frequency. The image
has been smoothed, as its purpose is only to show the general trends in the
energy-frequency parameter space. At low frequency (left), the lag tends to increase with energy.
At intermediate frequencies ($\sim 3.5\times10^{-4}$ Hz), the lag is negative around 2 keV. Now given
this rough idea of the energy-frequency dependence, we investigate it further.
\begin{figure}
\centering
 \includegraphics[width=200pt,clip ]{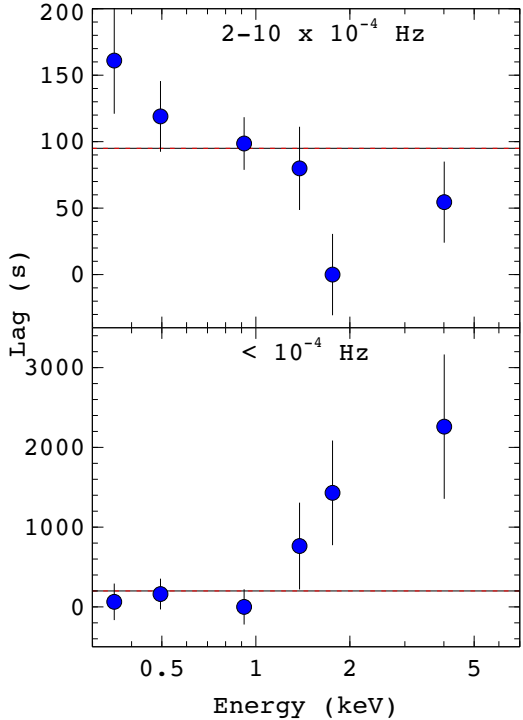}
\caption{Lag dependence on energy for the first observation of RE J1034+396 at high ($(0.2-1)\times10^{-3}$ Hz, top) and low
($<10^{-4}$ Hz, bottom) frequencies. Lags are calculated with respected to the entire energy band (0.3--10 keV),
then shifted so the minimum lag is zero. This shift implies the lag values in the top and bottom panels represents lags with respect to different bands.}
\label{fig:lag_energy1}
\end{figure}

Fig. \ref{fig:lag_energy1} shows the energy dependence of the lag for two frequencies ($<10^{-4}$
and $(0.2-1)\times 10^{-3}$ Hz). Light curves are extracted in six energy bands, defined to roughly have the same number of counts. The values of the lag have been shifted so that the minimum is at
zero. Because the lag is calculated taking the entire energy band as a reference, only the
difference between the bands and the shape are of importance. The shapes are clearly different in
the two frequencies, and roughly resemble those of 1H0707-495 (\citealt{2011MNRAS.412...59Z}). At low
frequencies, the lag increases with energy. This is the behaviour seen in several Galactic accreting
black holes (\citealt{1989Natur.342..773M}). Below 1 keV, the lag is more constant, indicating
the presence of a separate spectral component. At higher
frequencies, the lag dependence is reversed, and now decreases with energy, with a turn back
at $\sim 4$ keV.

The energy spectrum has at least two separate components (e.g. \citealt{2009MNRAS.394..250M}),
dominating the spectrum above and below 1 keV respectively. This appears to be consistent with the
low frequency lags. More interestingly however, there is a \emph{soft} lag between the soft ($< 1$
keV) and harder band ($>1$ keV), in a sense that the hard band \emph{leads} the variability
at $\sim 3.5\times10^{-4}$ Hz. This is strikingly similar to 1H0707-495 (\citealt{2011MNRAS.412...59Z}), and 
would indicate that the soft band ($< 1$ keV) is dominated by reprocessed emission, while the hard
band is dominated by a direct power-law component. The spectrum is investigated in detail in \mbox{Sec. \ref{sec:spec}}.

\begin{figure}
\centering
 \includegraphics[width=180pt,clip ]{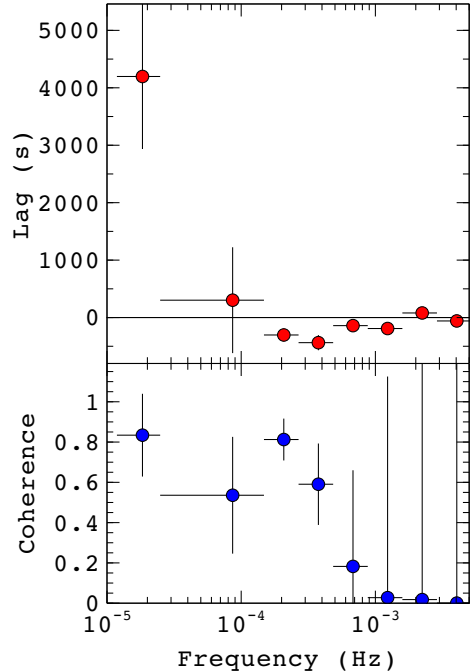}
\caption{Top: Lag as a function of frequency between 0.4--0.6 and 1.5--2.0 keV bands. Bottom: The corresponding estimated coherence function. The coherence was measured following \citealt{1997ApJ...474L..43V}.}
\label{fig:lag_coh_freq}
\end{figure}

\begin{figure}
\centering
 \includegraphics[width=220pt,clip ]{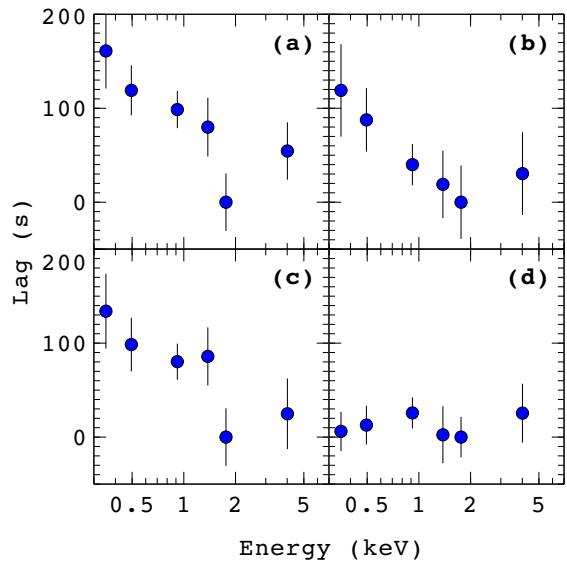}
\caption{Lag as a function of energy for the first XMM observation at $(0.2-1)\times 10^{-3}$ Hz showing the effect of pile-up on the lag. Panels a,b,c correspond to the lags measured when the source region is an annulus with inner radii of 0, 7 and 32 arcseconds respectively. Panel (d) shows the same plot from a simulated piled-up event file using {\sc scisim} where no lag is expected. }
\label{fig:lag_energy_pu}
\end{figure}

\subsection{Lag significance}
In order to check the lag significance, we first explore the coherence between the bands at the frequency of interest. The coherence function is a measure of the linear correlation between any two bands at the specified frequency (see \citealt{1997ApJ...474L..43V}). The soft lag is maximum between $\sim 0.5$ and $1.5-2$ keV, and we concentrate on measuring the coherence between these bands (0.4--0.6 and 1.5--2 keV). Fig. \ref{fig:lag_coh_freq} shows the estimated lag and coherence between the 0.4--0.6 and 1.5--2.0 keV energy bands. The lag is positive at low frequencies (hard lags) and negative at high frequencies (soft lags). The corresponding coherence shows a significant coherence between the bands below $\sim 8\times10^{-4}$ Hz. The coherence appears to drop slightly at $10^{-4}$ Hz. This is again similar to 1H0707-495, and could be related to a transition between two variability processes dominating the low and high frequencies.

The frequency of the soft lags in RE J1034+396 ($\sim3.5\times10^{-4}$ Hz) compared to those in
1H0707-495 ($\sim2\times10^{-3}$ Hz; \citealt{2011MNRAS.412...59Z}) reduces the effect of Poisson
noise on the lag signal. We checked this using Monte Carlo simulations of light
curves that have the same rms, inter-band lag and coherence as those observed. We followed the method of \cite{1995A&A...300..707T}. The details of the simulation are very similar to those in \cite{2010MNRAS.401.2419Z}. The lags in the simulation are recovered
and their RMS spread is consistent with the measured errors.

The first observation is highly piled-up. We repeated the lag calculations for several extraction
regions, excluding the central parts of the PSF (using annuli with inner radii of 0,7,14 and 32 arcsec).
This has very little effect on the shape of the lag-energy plot. The results are shown in Fig.
\ref{fig:lag_energy_pu}. The plots are for annuli of inner radius of 0 (a), 7 (b) and 32 (c)
arcsec. To make a further test, we used the XMM-\emph{Newton} {\sc scisim} packages to simulate a
whole piled-up observation. The input of the simulation is a point source with a flux of
$8\times10^{-12}$ ergs~cm$^{-2}$~s$^{-1}$ in the 0.3--10 keV band and a spectrum taken as a two
power-law component model fitted to the data. The simulated events file were reduced in the normal
procedure (using {\sc SAS v10.0.0}). The pile-up was of the same strength as in the real
observation. The lag-energy dependence of the simulated data is plotted in Fig.
\ref{fig:lag_energy_pu}-d. The lag is constant as expected, and pile-up does not effect the lag
measured at the frequencies of interest.

\section{Energy Spectrum}\label{sec:spec}
\subsection{Spectral fitting}
RE J1034+396 has one of the strongest soft excesses known. It was studied extensively in
\cite{2009MNRAS.394..250M}. Using constraints from the RMS spectra, a Comptonisation model was
favoured. Here, we explore the spectra from both observations, and include information from the lag
spectra. 

\begin{figure}
\centering
 \includegraphics[width=220pt,clip ]{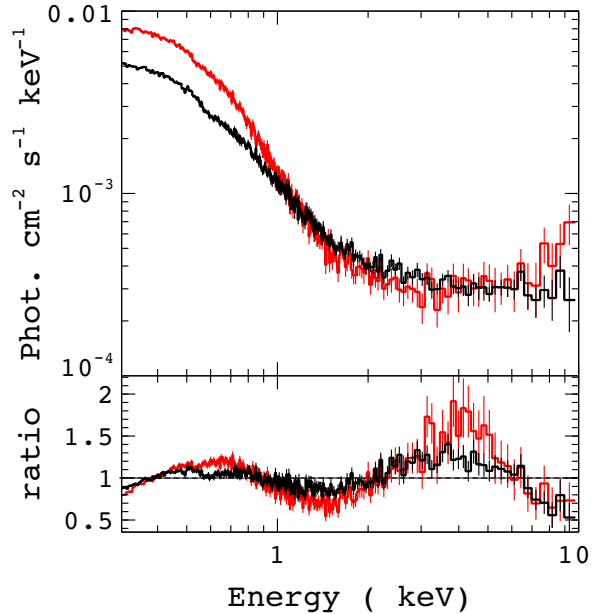}
\caption{The spectra from the two XMM observations. Top: Unfolded spectra produced in a model-independent way following Nowak et al (2005). The bottom panel shows the ratio of the spectra to a model consisting of a sum of two power-law models.}
\label{fig:unfolded_spec}
\end{figure}
Fig. \ref{fig:unfolded_spec} shows the spectra from the two observations. They are plotted as unfolded spectra (top) in a model independent way following \cite{2005ApJ...626.1006N}. This corrects for the detector effective area
and shows the main features in the spectrum. The bottom plot shows the ratio of the data to a double power-law model. All parameters of the model have been allowed to vary in the fit. The aim of these plots is to highlight the spectral changes and differences between the two observations. 
The spectrum not only appears to be pivoting around the 1 keV point indicating separate components above and below 1 keV, but also seem to have structural changes below 1 keV that could be related to atomic features in the spectrum of the soft band. This is possibly an indication that the soft component (below 1 keV) is not a smooth power-law-like component, but rather it has some atomic features and the difference between the observations is partly due to ionisation changes.

Atomic features in the spectrum (or changes thereof) can be due to absorption, reflection or a
combination of both. The smoothness of the spectrum itself requires the features to be smeared out
by high velocity shear in a wind or by relativistic broadening. Smoothing absorption features
requires unrealistically high velocities ($\sim c$, \citealt{2008MNRAS.386L...1S}), so a remaining
possibility is relativistic broadening of reflection emission, which is also motivated by the lag results. We used a model similar to that used
for 1H0707-495 (\citealt{2009Natur.459..540F,2010MNRAS.401.2419Z}) consisting of a power-law and
blurred reflection. We used the ionised reflection table model from \citealt{2005MNRAS.358..211R}
convolved with relativistic kernel {\tt kdblur} (\citealt{1991ApJ...376...90L}). Galactic column in
the direction of the source is $1.52\times10^{20}$ cm$^{-2}$ (\citealt{1992ApJS...79...77S}, modelled in Xspec with \texttt{tbabs}, \citealt{2000ApJ...542..914W}). To obtain more constraints at low energies, we also included an earlier
observation from the ROSAT PSPC archives that extends down to 0.1 keV. This provides
better constraints on the accretion disc thermal emission peaking in the EUV and expected to
contribute to soft X-rays (\citealt{1995MNRAS.276...20P}).
\begin{figure}
\centering
 \includegraphics[width=240pt,clip ]{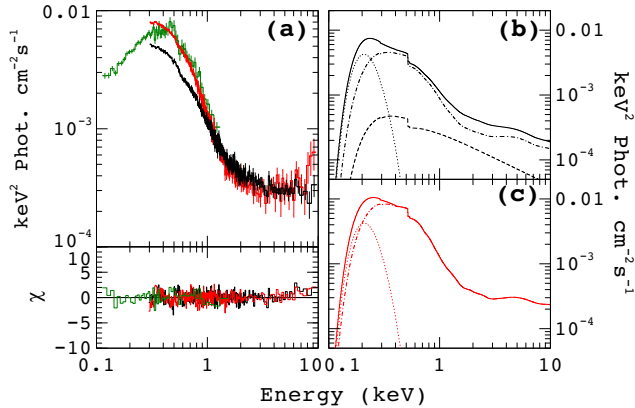}
\caption{\texttt{(a)} The spectrum of RE J1034+396 from the two XMM observations (black and red), and from the ROSAT PSPC (green) unfolded following Nowak et al. (2005). The bottom panel is a plot of the fit residuals in terms of sigma. \texttt{(b)}: The best fit model for the first XMM observation. The model consists of a low temperature blackbody, a power-law and a blurred reflection. \texttt{(c)}: Similar to \texttt{(b)} for the second XMM observation. The colours are similar to those in Fig. \ref{fig:unfolded_spec}}
\label{fig:spec}
\end{figure}

For initial modelling, we fitted the XMM-Newton data with a model consisting of a power-law plus blurred reflection. This simple model did not give a very good fit. The strongest residuals were an edge-like feature at $\sim 0.5$ keV. 
Adding an edge to the model improved the fit significantly with the edge having an energy of 0.49 keV. This corresponds to oxygen photoelectric absorption at the source redshift ($z=0.042$). The fact that it is variable indicate a change in the line-of-sight column density at the host galaxy. To model this, we included a photoelectric absorption component with variable abundance (\texttt{tbvarabs}). The redshift was fixed at that of the host galaxy. This improved the fit significantly ($\Delta \chi^2 > 100$ per extra degree of freedom). To obtain further improvements, we also added data from the ROSAT PSPC, so the model now extends down to 0.1 keV, which helps constrain the thermal emission from the accretion disc, and any contribution it may have in the XMM band.

Fig. \ref{fig:spec}-(a) shows the spectrum from the two XMM observation (black and red) and the
ROSAT observation (green). The spectra are again unfolded in a model-independent manner (\citealt{2005ApJ...626.1006N}). The models are plotted on the right for the first (top) and second (bottom)
XMM observations. The model provides a good fit
to the data ($\chi^2=698$ for 611 degrees of freedom). The blackbody has little contribution in the
XMM band and mainly contribute to energies less than 0.3 keV, and it is in
full agreement with multi-wavelength fits using optical to soft X-rays data
(\citealt{1995MNRAS.276...20P}), where there is no strong big blue bump in the optical/UV but the source has a strong emission in the EUV/Soft X-ray indicating hot accretion disc emission. The disc component in our case is mainly constrained by the ROSAT PSPC data,
and remains constant between the three observations. The best fitting parameters for the whole
model are presented in Table \ref{tab:fit}.
\begin{table}
\begin{center}
\label{tab:fit}
 \begin{tabular}{@{}ll@{}}
Parameter			& Value\\
\hline \\
$\Gamma_{\rm power-law}$			& $2.65\pm0.03$\\
$r_{\rm in}$			& $1.90^{+0.11}_{-0.12}$ ($r_g$)\\
$q$			& $6.24\pm0.5$ \\
$\xi_1$, $\xi_2$, $\xi_3$	& $715^{+25}_{-53}$,
$346^{+93}_{-80}$,$184^{+60}_{-7}$ (erg cm s$^{-1}$)\\
\hline
\end{tabular}
\caption{The fit parameters for the model shown in Fig. \ref{fig:spec}. $r_{\rm in}$ is the inner radius of the reflecting region. $q$ is the emissivity index of the disc, modelled as $\propto r^{-q}$. $\xi$ is the ionisation parameter.}
\end{center}
\end{table}

\begin{figure}
\centering
 \includegraphics[width=200pt,clip ]{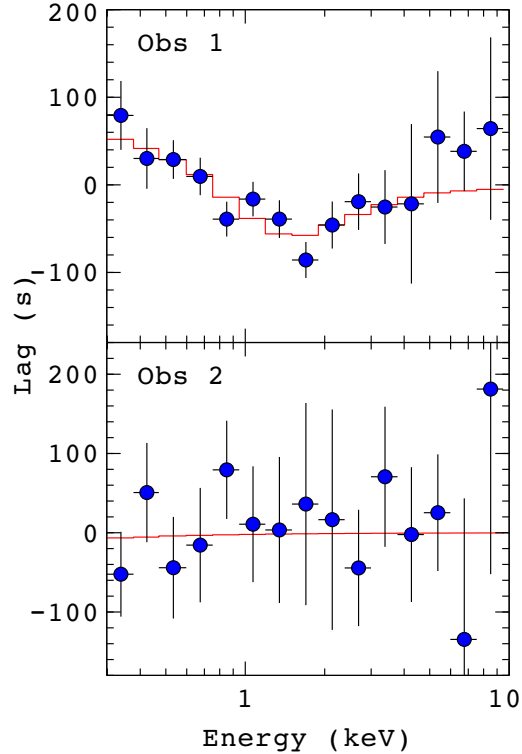}
\caption{Lag as a function of energy for the first (Top) and second observation (Bottom). The whole 0.3--10 keV band is used as a reference (removing the current band each time). The red line is a model constructed by taking the ratio of the power-law component in the best fit model to the total (power-law + reflection). The lines represent the best fit model. More energy bands are used here compared to Fig. \ref{fig:lag_energy1} to enable better energy resolution. The energy bands are selected in equal steps in log space.}
\label{fig:lag_energy_model}
\end{figure}

\subsection{Modelling the lag}
The soft lag discussed in Sec. \ref{sec:time_lag} should, when interepreted correctly, be related
to the energy spectrum. The spectral model (Fig. \ref{fig:spec}) show that for the first
observation, the relative contribution of the reflection compared to the direct power-law initially
decreases with energy, it reaches a minimum at $\sim 2$ keV, and increases again. This is precisely
the behaviour of the lag in Fig. \ref{fig:lag_energy1}. To show this more clearly, we plot in Fig.
\ref{fig:lag_energy_model} (top) the lag-energy dependence, along with a model representing the
ratio of the reflection component to the total spectrum (power-law + reflection) shown in Fig. \ref{fig:spec}. The constant offset
was added to the model to account for the reference, and multiplied by a factor representing the
intrinsic lag between the two components. The latter is found to be $296\pm80$ seconds.

The two XMM observations are different spectrally and in the light curve properties. The second
observation is reflection-dominated, where the direct power-law is not observed. Several AGN have
been observed sometimes to be in a reflection-dominated state, particularly at low fluxes (e.g.
\citealt{2008MNRAS.391.2003Z}). Although the second observation is shorter, we calculated the
energy-dependent time lag and show it also in Fig. \ref{fig:lag_energy_model} (bottom). The
data do not show the lags seen in the first observation and are consistent with a constant. This
is exactly what is expected, as the direct emission is not seen, and the lag now is measured
between different parts of the reflection component itself. Despite the high uncertainties in the second observation, the model fitted to the first can be ruled out in the second observation at around 95 percent confidence.

\section{Discussion}
We have shown that there is a soft lag between the hard and soft components in the light curves of
RE J1034+396. The simplest interpretation is that the soft component ($< 1$ keV) is
reprocessed emission produced in a partially ionised reflector. This object is very similar to 1H0707-495
in its lag properties, and its spectrum can be well modelled with a reflection model dominating the
soft band. The best fitting spectral model matches very well the lag-energy spectrum.

The smoothness of the soft excess requires the sharp reflection features to be smeared out by strong relativistic effects,
implying that the emission region is very close to the black hole. Spectral fitting indicate the inner radius of the emission region to be $<3$ gravitational radii, with an emissivity index of $\sim 6$, slightly flatter than that of the 1H0707-495 (\citealt{2010MNRAS.401.2419Z}), indicating a slightly extended emission region. Thermal emission from the accretion disc also contribute to the soft X-ray band (0.1-0.4 keV). The X-ray and EUV flux is unusually higher than the flux in the optical and UV, indicating that the big blue bump, due to the accretion disc, is shifted to higher energies (\citealt{1995MNRAS.276...20P}), and a small contribution is expected in the XMM PN band.

A low variability, low temperature Comptonised emission can reproduce the soft excess in this object (\citealt{2009MNRAS.394..250M}). However, it would be difficult, in such a model, explaining the temperature of the seed photons, which appears constant, independent of the black hole mass. It is also interesting to explore the long term spectral changes. The differences between the ROSAT and XMM data (\ref{fig:spec}-a) indicate that the soft band is indeed dominated by broad spectral features that change with time, possibly due to geometry and/or ionisation changes. Furthermore, the detection of a soft lag where the soft band (0.5-1 keV) lags the medium band (1-4 keV), with an energy dependence that has a minimum around 2 keV, also strongly supports the interpretation of the soft emission being due to reprocessing.

It is interesting to explore the measured lag quantitatively too. The lag in 1H0707-495, after correcting for the contribution of individual components, is $\sim 70 $ seconds (\citealt{2011MNRAS.412...59Z}) at a temporal frequency of $1-2\times10^{-3}$ Hz. In RE J1034+396, the measured lag is $\sim 290$ seconds at a frequency of $\sim 3-4\times10^{-4}$ Hz. There is a factor of $\sim 4$ difference in the lag and the time-scale of the lag, indicating possibly a factor of 4 difference in mass. The frequency at which the lag is measured corresponds to the viscous or thermal time-scales at the radius the emission, while the lag is a measure of the light travel time between the illuminating source and the reflector. Although the exact numbers depend on the exact unknown geometry, the lags are consistent with mass estimates of $1-2\times10^6$ M$_{\odot}$ and $4-8\times10^6$ M$_{\odot}$ for 1H0707-495 and RE J1034+396 respectively.

RE J1034+396 is the only Seyfert galaxy known to show a quasi-periodic oscillation (QPO)  in its light curve (\citealt{2008Natur.455..369G,2009MNRAS.394..250M}). The QPO is present in the first XMM observation but not in the second. The spectra of the source in the two observations appear to be different (Fig. \ref{fig:spec}). Our spectral fitting has shown that the spectrum in the second observation is reflection-dominated. In the first observation however, the fitting requires the presence of a power-law component that contributes significantly around 2 keV. This indicates a possible relation between the QPO and  the power-law component in the energy spectrum. This is also supported by the shape of the rms spectrum of the QPO (Fig. 2 in \citealt{2009MNRAS.394..250M}), where the rms power in the QPO peaks around 2 keV. 
\cite{2010ApJ...718..551M} attributed the QPO to an orbiting cloud close to the black hole, producing mainly an absorption edge at $\sim 8$ keV. However, the apparent edge change is more likely to be due to continuum change and not a change in the atomic features. The shape of the RMS spectrum rules out this possibility.
Linking the QPO to the power-law component is in fact not surprising given what is known about QPOs in Galactic black holes (GBH), where the QPO is associated with the hard spectral tail and not the thermal disc component (e.g. see \citealt{2006ARA&A..44...49R}). This however, does not necessarily mean that observing a power-law component implies detecting a QPO. 

QPOs in GBH come generally in two types: low- and high-frequency QPOs and are observed in pairs with a frequency ratio of 3:2 ( \citealt{2006ARA&A..44...49R} and references therein). Identifying the equivalent of the QPO in RE J1034+396 is not clear, but taking its mass to be a few $\times10^6$ M$_{\odot}$, makes the QPO more likely to be the equivalent to the high-frequency type (HFQPO, \citealt{2008Natur.455..369G,2009MNRAS.394..250M}). A simple relation between mass and the fundamental frequency was suggested for GBH of the form: $M=931 f_0^{-1}M_{\odot}$, where $f_0$ is the fundamental frequency of the QPO (\citealt{2006ARA&A..44...49R}). If we assume it applies also to AGN, the mass of RE J1034+396 would be in the range $6-10\times10^6$ M$_{\odot}$ (depending on which harmonics the observations corresponds to), which is very consistent with the estimates from the lag.

The lag in RE J1034 extends over a wide frequency range, similar to 1H0707-495. The frequency of the QPO (a narrow frequency band by definition) is smaller than the frequency where the lag is maximum (Fig. \ref{fig:en_fq} and \ref{fig:lag_coh_freq}). The accretion disc is reflecting the random stochastic signal of the illuminating source, which is this case happens have a QPO, but generally it does not. 
The difference in frequency between the QPO and the lag is possibly an indication that the corona producing the QPO is slightly extended, while the compact reflection from the disc is emitted in a very small region and dominated by light-bending effects. This can also explain the absence of the QPO in the reflection component, only the inner parts of the variable power-law illuminate the inner regions of the disc and not the regions producing the QPO.

\section*{Acknowledgements}
AZ thanks the Cambridge Overseas Trust and STFC.
ACF thanks the Royal Society for support. The authors thank the anonymous referee for their
comments and useful discussion.

\bibliography{bibliography}
\label{lastpage}

\end{document}